\documentstyle[preprint,prd,aps,floats,epsfig]{revtex}

\begin{document}
%\input epsf

%\renewcommand{\topfraction}{0.8}

%@twocolumnfalse\endcsname

\title{Chaos and preheating}

\author{S. E. Jor\'as$^{1,}\thanks{Email address: joras@if.ufrj.br}$
 and
V. H. C\'{a}rdenas$^{2,}\thanks{Email address:
victor.cardenas.v@mail.ucv.cl}$}

\address{$^1$Instituto de F\'{\i}sica --- UFRJ, Caixa Postal 68528,
 21945-970
Rio de Janeiro, RJ --- Brazil }

\address{$^2$Instituto de F\'{\i}sica, Universidad Cat\'{o}lica de
Valpara\'{\i}so, Casilla 4059, \\ Valpara\'{\i}so, CHILE}

\date{\today}

\maketitle

\begin{abstract}
We study the semiclassical particle creation in the preheating phase
 after  inflation. We work in the long-wavelength limit, in which all
fields  are  considered homogeneous. The particle creation is shown
to be  intrinsically connected to the existence of  chaos in the
system.

\end{abstract}


%\vskip2pc]

%%%%%%%%%%%%%%%%%%%%%%%%%%%%%%%%%%%%%%%%%%%%%%%%%%%%%%%%%%%%

\section{Introduction}

    One of the most successful ideas to describe the early universe
 is inflation \cite{infmod}. Its main problem may be how to leave
behind  an exponential scale factor and reach a radiation-dominated
universe in  thermal equilibrium. A possible way to solve this
question is to turn on the interaction of the inflaton with another
scalar field, typically  radiation, when the former reaches the
bottom of its potential. Initially, this interaction was introduced
via a somewhat {\it ad hoc} dissipation  term  characterizing the
so-called {\it reheating} process \cite{reheat}.  Later on, it was
assumed a quadratic coupling between two scalar fields, which allowed
for the transfer of energy from the inflaton to the radiation by
parametric resonance: this is the {\it preheating} phase
\cite{robert,kofman,boyan}. This model yields a much more effective
amplification of particular modes  of the radiation, at the same time
raising some questions about the thermal  state of the universe after
the process is over \cite{feldkof,turb}.

    In the usual approach to preheating models, one assumes a fixed
evolution for the inflaton, and calculates the equation for a given
 mode of the coupled field $\chi$, which depends on the particular
inflaton  potential chosen. In any case, the inflaton behaves as an
infinite reservoir of  energy, driving the exponential amplification
\begin{equation}
\chi_k\propto \exp(\mu_k t)
\end{equation}
characterized by the Floquet exponent $\mu_k$, of particular modes
 $k$ of the radiation field $\chi$ indefinitely.

    In this paper we consider both fields as a coupled
system  in a given background and investigate their properties; of
 particular interest are the issues of how chaotic the dynamics is
and what is  the relation between the resonance effects and its
chaotic character.

Indeed, as we will shortly show, chaos arises precisely at the time
when exponential amplification  occurs. We further propose a
relationship between the metric entropy, usually  defined for chaotic
systems, and the entropy corresponding to the particles  produced
after inflation by parametric resonance.

    As will be seen below, in our simple model there is no interaction
 among different modes; thus, as in any system with a finite number
of  degrees of freedom, the energy will keep oscillating from one
field to the other indefinitely. In practice, we expect that it
will eventually be transferred into other fields. We conjecture
this can be taken into account  introducing by hand a viscosity
term in the equation of motion for the radiation --- this
 analysis  will be accomplished elsewhere
\cite{future}.

Note that there are fundamental differences between this work and
previous papers in similar subjects. In Ref. \cite{corlev},  the
 existence of chaos on the evolution of the {\it scale factor} was
investigated;  here, the background evolution is given {\it a
priori}. We stress that we are  interested  in the beginning of the
preheating phase, when backreaction ---  the  effect of the amplified
field on the evolution of the scale factor --- is not  strong  enough
yet. Actually, we will consider a {\it static} universe, since the
 parametric  resonance  happens in a much shorter time scale than the
expansion of a radiation-dominated universe  \cite{robert}.
 The authors of Ref. \cite{EK-M} also discussed the chaotic behavior
 in the case of two-field inflation, but they used a symmetry
breaking  potential and investigated the enhancement on the
production of topological  defects. We have chosen a single-well
potential for the sake of simplicity; a double-well potential would
eclipse our point somehow. Ref. \cite  {feldkof} studies the approach
to equilibrium for a couple of different  potentials, but being
interested in the turbulent phase right {\it after} preheating,  the
 authors introduce a ``normalized distance'' in the phase space,
according to  which chaos sets in {\it after} the preheating is over.
Here we apply the  usual recipe \cite{wolf,holden} for computing the
largest Lyapunov  exponent, as explained below. As we mentioned
above, we will show that parametric  resonance and chaos seem to be
fundamentally related.

        In spite of the aforementioned assumptions, our model is able
 to grasp  important qualitative features such as the energy
threshold above which the  dynamics  becomes chaotic. The next
section illustrates the relation between the  Lyapunov (LE) and
the Floquet (FE) exponents for two well known problems: the
parametrically  excited pendulum and the Mathieu equation. In section
\ref{model} we  introduce the  analogous question for the coupled
scalar fields in cosmology, followed by a  brief  description of the
classical chaotic properties of the system. In section \ref{semi}, we
follow a simple and transparent  method  \cite{PC}, consistent with
Heisenberg's principle, to obtain a semiclassical  approximation
 for our model. We then study the onset of chaos on the effective
 equations of motion  which describe the particle production process.
Then we discuss the  relationship  between  ({\it a priori})
different entropy definitions in section  \ref{entropy}.

%%%%%%%%%%%%%%%%%%%%%%%%%%%%%%%%%%%%%%%%%%%%%%%%%%%%%%%%%%%%%%%%%%%%%
%%%%%

\section{Parametric Resonance}

    A positive Lyapunov Exponent (LE) is the main characteristic of
 chaotic motion,  as it indicates a strong sensitivity to small
changes in the initial  conditions. The LE measures the mean
separation of two initially  neighboring  trajectories in the phase
space, in logarithmic scale:
\begin{equation}
\lambda_i=\lim_{t\rightarrow\infty}
\left\{\frac{1}{t} \ln\left[ \frac{L_i(t)}{L_i(0)}\right] \right\}
\end{equation}
Because such separation soon approaches the size of the attractor, a
 naive  computation using the expression above would fail to detect
the  local rates of  expansion. Thus the distance between the
trajectories must be  periodically  normalized; the LE will then be
the average of the exponential rates  obtained  this way.

    The parametrically excited pendulum
\begin{equation}
\ddot{\theta} + 2\eta \; \dot\theta + [1 + p \cos(\omega t)]     \sin
 (\theta) =0
\label{pend}
\end{equation}
is known to be chaotic \cite{pendulum}. Indeed, one can calculate its
 largest  Lyapunov exponent and find a positive quantity. On the
other hand,  the Floquet  theorem assures that the solution of
eq.~({\ref{pend}) is given by
\begin{equation}
\theta(t) = f_P(t)\exp (\pm \mu \;t)
\end{equation}
where $\mu$ is the Floquet exponent. It is easy to see that both
 exponents must actually be the same, since the only difference
between them is  the normalization procedure in the calculation of
the LE. In the  parametrically excited pendulum such procedure is
not needed, since the independent  variable $\theta$ is cyclic: it
is never too large. The same reasoning applies  to the study of
metric perturbations in a chaotic background: the rate of  growth
of perturbations --- valid only in the {\bf linear} regime --- is
given
 by the LE, as shown in Ref.~\cite{zibin}.

        The relation between both exponents is also clearly seen in
 the  typical parametric resonance phenomenon, described by the
Mathieu  equation
\begin{equation}
\ddot R (t)=- [ \Omega^2 + g^2 x^2(t) ]R(t)\;
\label{mathieu}
\end{equation}
where $x(t)=\sin (\omega t)$. We used $\omega=10$,
$\Omega=\sqrt{4/5}\omega$, and $g=2\omega/\sqrt{10}$. The FE for the
above equation can be  exactly calculated \cite{floquet}, and for the
used values one obtains  $\mu=0.5$. For the sake of completeness and
as a test of our numerical code, we  calculated it by plotting
$(1/2t) \ln(n_R) ~{\it versus}~  t$, where $
n_R=\Omega/2(|R(t)|^2+|\dot R(t)|^2/\Omega^2)$, can be interpreted as
the energy. Fig. \ref{parametric} shows that the LE and both the
calculated and theoretical values for the FE converge to the same
value.

    If the phase space is not limited --- as in the case of exact
 parametric resonance described by the Mathieu equation
(\ref{mathieu}) --- one  cannot rely on  the LE to tell if the system
is chaotic or not \cite{ott}.  Nevertheless, actual  physical systems
will have only a finite amount of energy  available. Then, the
available phase space will be finite and the LE criterium for the
existence of  chaos will  hold.

%%%%%%%%%%%%%%%%%%%%%%%%%%%%%%%%%%%%%%%%%%%%%%%%%%%%%%%%%%%%%%

\section{Coupled fields}
\label{model}

 Most of the work in preheating has been made asuming a biquadratic
 coupling  between the inflaton $\phi$ and the secondary field
$\chi$:
\begin{equation}
V_{\rm int}=g^2 \phi^2 \chi^2
\label{potint}
\end{equation}

We assume a flat FRW universe background whose line element is
 written as
\begin{eqnarray}
ds^{2} &=& dt^{2}-a^{2}(t)d{\bf x}^{2},  \nonumber \\
&=& a(\eta) (d\eta^2 - d{\bf x}^{2})  \label{ec2}
\end{eqnarray}
where $a(\eta)$ is the scale factor in conformal time. In this metric
 the equations of motions are
\begin{equation}
Y^{\prime \prime }(\eta )+\Omega ^{2}(\eta )Y(\eta )=0,  \label{ec6}
\end{equation}
\begin{equation}
X^{\prime \prime }(\eta )+\omega ^{2}(\eta )X(\eta )=0,  \label{ec7}
\end{equation}
where $\Omega ^{2}=a^{2}M^{2}+g^{2}X^{2}-a^{\prime \prime }/a$ and
 $\omega ^{2}=a^{2}m^{2}+g^{2}Y^{2}-a^{\prime \prime }/a$, $X\equiv
a\chi $  and $Y\equiv a\phi$. If we restrict ourselves to the
beginning of the  preheating phase, the backreaction can be safely
neglected. In this case, the evolution  of the universe is nearly
a radiation dominated phase, with $a(\eta)=\eta/2$  and thus
$a^{\prime\prime}/a=0$. Actually, for the sake of simplicity, we
rely on the much faster dynamics of preheating to  assume a static
universe, and, for convenience we adopt $a(\eta)=1$ (and thus
$t\equiv \eta$). By doing so we are neglecting the time dependence
of the  instability bands; we expect the qualitative aspects of
the results presented  here to remain unaltered in a more refined
analysis \cite{future}. We work in the  long-wavelength limit and
assume all fields as homogeneous. Indeed, the main  contribution
to
 the  parametric resonance is due to the zero mode\cite{boyan2}.

    At the classical level, the system of Eqs.(\ref{ec6},\ref{ec7})
 describes a well know chaotic system \cite{chaos}. Indeed, if we use
the  standard technique \cite{wolf,holden} we can calculate the
largest LE for two  different trajectories. Figure \ref{le2} shows
two such plots, one of which is non-chaotic.
Since this system presents a finite number of degrees of freedom, the
 energy will oscillate back and forth between them. Thus a given
variable  will increase exponentially only during the energy
transfer. The naive way  to determine the FE is to look at the  plot
of $\frac{1}{2\eta}\ln[Y^2(\eta)/Y^2(0)]$ but, because of the
oscillation just mentioned, we would have to reset the time  variable
at the beginning of each phase, just as for the chaotic pendulum.
 Instead, we decided to plot $\frac{1}{2}\ln [Y^2(\eta)/Y^2(0)]
\mbox{\it versus }  \eta$ and looked at the angular coefficients of
the straight lines. Figure  \ref{fe2} shows such plot for the same
two initial conditions used in the  previous figure.

The last two figures seem to show that the LE and the FE are
 numerically equal. For the parameters chosen the resonance period,
although clearly  noticeable, is too short for a statistical
analysis that would allow us for a  measure of the small
discrepancy between the obtained values for the FE and LE.

%%%%%%%%%%%%%%%%%%%%%%%%%%%%%%%%%%%%%%%%%%%%%%%%%%%%%%%%%%%%%%%%%%%%%
%%%%%%%%

\subsection{Energy threshold}

    We were able to numerically determine the existence of an energy
threshold below which there is no amplification of the  secondary
 field. Figures \ref{thL} and \ref{thF} show a plot of the LE and FE,
 respectively, for ten randomly chosen initial conditions. One can
see that the  trajectories with energies below a critical value
are restricted to a certain region,  mainly below the horizontal
axis of the latter graph, while the ones above  that limit
oscillate with a much larger amplitude. Note that the trajectory
presenting a non-vanishing LE corresponds to the upper curve in
Fig. \ref{thF},
 precisely because it indicates a non vanishing net value for its
angular  coefficient.
An exact calculation of the energy threshold can be done by plotting
 the fraction of the phase space covered with invariant tori
 \cite{gutzwiller}; this analysis will be presented elsewhere
\cite{future}.

    In order to explore this statement, we shall study the stability
properties of the potential $V(X,Y)$. From Eq.(\ref{potint}) we know
 the potential (both mass and interaction terms) is
\begin{equation}
V(X,Y)=\frac{1}{2}(M^{2}Y^{2}+m^{2}X^{2}+g^{2}Y^{2}X^{2}),
\label{potxy}
\end{equation}
Using the Toda-Brumer-Duff test for instabilities \cite{TBD}, we find
 that the system develops instabilities for energies larger than
\begin{equation}
E^{\ast }\simeq \frac{m^{2}M^{2}}{g^{2}}, \label{estar}
\end{equation}
which indeed lies between the two energy ranges presented in the two
 previous plots. We stress that this result does not mean that chaos
must not happen  for energies below that critical value.

%%%%%%%%%%%%%%%%%%%%%%%%%%%%%%%%%%%%%%%%%%%%%%%%%%%%%%%%%%%%%%%%%%%%%
%%

\section{Particle production}
\label{semi}

As we saw in the last section, even the classical-field-theory
version of our model exhibits chaos. Note that we do not have the
right to use words like `preheating' and `particle creation'
because they only have meaning in the quantum version of the
model. In this section, we use the method presented in
Ref.\cite{PC} in order to take into account the quantum effects.
We will show that most of the results of the last section can be
used to describe the semiclassical system. We called this system
semiclassical because we are assuming that $Y(\eta )$ is a {\it
classical} field and $X(\eta )$ is treated as a {\it quantum}
operator. We have to stress here that this procedure is equivalent
to take the large $N$ approximation (see Ref.\cite{QC}).

Let us expand the quantum field $X$ in the Heisenberg
representation
\begin{equation}
X(\eta )=f(\eta )\,a+f^{\ast }(\eta )\,a^{\dagger },  \label{2ec1}
\end{equation}
where $a$ and $a^{\dagger }$ are annihilation and creation
operators satisfying the standard commutator $\left[ a,a^{\dagger
}\right] =1$. From this result, the mode function $f(\eta )$ has
to satisfy the Wronskian condition
\begin{equation}
f^{\prime \ast }f-f^{\ast }f^{\prime }=-i,  \label{2ec2}
\end{equation}
and from Eq.(\ref{ec7})
\begin{equation}
f^{\prime \prime }(\eta )+\omega ^{2}(\eta )f(\eta )=0.
\label{2ec3}
\end{equation}
In order to satisfy Eq.(\ref{2ec2}) and Eq.(\ref{2ec3}) we use the
Ansatz
\begin{equation}
f(\eta )=\frac{1}{\sqrt{2W(\eta )}}\exp \left( -i\int^{\eta }d\eta
^{\prime } \; W(\eta ^{\prime })\right) .  \label{2ec4}
\end{equation}
Choosing the vacuum $\left| 0\right\rangle $ of the number
operator $ n=a a^{\dagger }$, defined by $a\left| 0\right\rangle
=0$, to compute averages $\left\langle ...\right\rangle $, we
obtain an effective Lagrangian $L_{eff}=\left\langle
L\right\rangle $
\begin{equation}
L_{eff}=\frac{a^{2}}{2}\left[ Y^{\prime 2}+R^{\prime
2}-\frac{1}{2R^{2}} -a^{2}m^{2}Y^{2}-\omega ^{2}R^{2}\right] ,
\label{2ec5}
\end{equation}
where we have defined a new field $R^{2}(\eta )=1/2W(\eta )$. In
this case, the equations of motion are
\begin{equation}
Y^{\prime \prime }(\eta )+\Omega ^{2}(\eta )Y(\eta )=0,
\label{2ec6}
\end{equation}
\begin{equation}
R^{\prime \prime }(\eta )+\omega ^{2}(\eta )R(\eta
)-\frac{1}{4R^{3}(\eta )} =0, \label{2ec7}
\end{equation}
where the averaging process redefined $\Omega ^{2}(\eta
)=a^{2}M^{2}+g^{2}R^{2}-a^{\prime \prime }/a$. Because
$R^{2}=\left\langle X^{2}\right\rangle $ the centrifugal term,
$1/4R^{3}$, keeps the quantum expectation value away from zero,
consistently with the Heisenberg's uncertainty principle and
splits in two the phase space of $R(\eta )$. Note that now the
amplification of $R$ can actually be interpreted as particle
production if also the number
\begin{equation}
n_{R}=\frac{\omega }{2}\left[ \frac{\dot{R}^{2}}{2\omega
}+R^{2}\right] - \frac{1}{2},  \label{number}
\end{equation}
grows.

The question is: Is the system described by Eqs.
(\ref{2ec6},\ref{2ec7}) chaotic?. As we have described in section
II, the main feature of chaotic motion is a positive Lyapunov
exponent, which shows a sensitivity to small changes in the
initial conditions. Another way to reveal chaotic behavior is the
study of Poincar\'e sections. A Poincar\'e section or a surface of
section, is a two dimensional map of the phase space obtained by
intercepting the hamiltonian flow at a fixed position during the
motion. For integrable systems, the map is a collection of lines
and regions of stability. As soon as the system becomes chaotic,
the lines become distorted and the stability regions disappeared.
Our investigation of this issue for the system in Eqs.
(\ref{2ec6}, \ref{2ec7}) shows that it is indeed chaotic.

Our task is made somewhat easier if we separate the analysis
between $R\gg 1$ and $R\sim 1$ regions. When $R\gg 1$, the system
(\ref{2ec6}, \ref{2ec7}) behaves similarly as described in the
last section, e.i. Eqs. (\ref{ec6}, \ref{ec7}) but constrained to
one of the two halves of the phase space. This last statement is
not rigorously required, because we can leave the system evolving
through the barrier at $R=0$ without affecting the general
properties of the chaotic system, e.g.: both have the same LE.

In the small $R$ region, where the centrifugal term is important,
we find an interesting saturation effect. Because we assume that
initially the fluctuation $R$ is not big (its minimum is around
$1/\sqrt{2\omega }$), there is a transient period, where although
a resonance condition is fulfilled, the $R$ field grows until the
nonlinear term breaks the resonant tuning. This fact is indeed
connected with the existence of a critical energy under which the
system does not amplify the field, as we see later. Unfortunately,
we were not able to calculate the LE here because of the very
phenomenon we are studying, the parametric resonance itself. As
the $R$ oscillation amplitude is increased, $R$ gets closer and
closer to the origin, and then the centrifugal barrier just
explodes. Therefore, one cannot follow the evolution of the system
for a time long enough to allow the graph to reach the constant
value which would correspond to the largest LE.

Nevertheless, we can show the existence of chaos, and even
conjecture about a threshold, by plotting Poincar\'e sections
valid for both regions ($R\gg 1$ and $R\sim 1$) for different
values of energy, as shown in figure \ref{poincare}.

By using the Toda-Brumer-Duff instability condition we can
estimate a critical energy under which the system does not amplify
efficiently the field $R$. For the case $R\gg 1$ the result is
 Eq. (\ref{estar}). In Fig. 6 we can see the gradual
destruction of the tori as the energy increases. For the values in
the numerical example, $M=10, m=1, g=1$ we obtain $E^{\star} \sim
100$. The upper panel shows the $E=50$ case well inside the
integrable region, shows clearly continuous lines with a single
stability region, the central one. The central panel shows the
$E=150$ case, beyond our estimation for stability, where is
possible to see that the original continuous exterior lines have
disappeared, and in their place there is a band of scattered
points. The case $E=200$ shows the same effect even more
dramatically.

%
%
%
%

%%%%%%%%%%%%%%%%%%%%%%%%%%%%%%%%%%%%%%%%%%%%%%%%%%%%%%%%%%%%%%%%%%%%%
%%%%%%

\section{Entropy production}
\label{entropy}

In the previous sections we have shown evidence for a relation
 between classical chaos and particle production via a correspondence
between  the FE, which characterizes exponential growth during
preheating, and the maximal positive LE for the associated chaotic
system.  In this section we discuss further consequences of this
relation.

On one hand, we know that for a chaotic dynamical system we can
 define a {\it metric} or {\it Kolmogorov entropy} ${\cal K}$
\cite{holden} in  terms of the LEs $\lambda_i$ as
\begin{equation}
{\cal K} \leq \sum_{ \{\lambda_i\}>0 }\lambda_i\;,  \label{kolmo}
\end{equation}
--- where the equality holds for {\it typical} Hamiltonian systems
 \cite{ott} --- which gives the rate of change of the available
information.  Local trajectories get stretched in the direction in
which the eigenvalues  $\lambda_i$ are positive, and get
compressed in the directions in which they are  negative. If there
are no positive LEs then there is no change in the amount of
information available, and the Kolmogorov entropy vanishes. In our
problem, there
 are at least two directions in phase space with comparable, actually
equal,  LE: $R$ and $\dot R$. Thus
\begin{equation}
{\cal K} \leq 2\lambda\;.
\end{equation}
On the other hand, the process of particle creation can also be
 described as a period of entropy production. The problem is then try
to find a  formula for entropy valid during the transfer of energy
between the  oscillators, i. e., in a non-equilibrium system. A
lot of work has been done in this  context (see Ref.\cite{hu} for
a review), each one arguing in favor of a  different definition
for the entropy and its physical reasoning as such. In  spite of
their different conceptual foundations, all of them agree that, in
the high squeezing limit,
\begin{equation}
S\approx \ln (n_k)\;,
\end{equation}
where $n_k$ is the number of particles in the mode $k$. In our case,
 it is given by $n_k \propto \exp (2\mu t)$, and then
\begin{equation}
S\approx 2\mu t\;,
\end{equation}
showing the equivalence between the Lyapunov and the Floquet
 exponents once more.

%%%%%%%%%%%%%%%%%%%%%%%%%%%%%%%%%%%%%%%%%%%%%%%%%%%%%%%%%%%%%%%%%%%%%
%%%%%%

\section{Discussion}

We notice that the reheating process after inflation, in
particular the initial stage called preheating, could be driven by
a different dynamical behavior, more complicated than the
currently believed parametric resonant picture. We investigated
the preheating phase by using a system comprised by two
interacting background fields. We showed that the study of this
model simplifies to that of two coupled harmonic oscillators. Our
main results suggest a strong correspondence between the
parametric resonance phenomenon and the chaotic properties of the
system, namely the numerical equivalence of its Floquet and
Lyapunov exponents.

We study the onset of chaos on the effective equations of motion
which describes the particle production process. Nevertheless, we
can only say the system is {\it strongly dependent on the initial
conditions}, since the LE are not unambiguously characterize chaos
in General Relativity. We can talk about chaos with additional
information, for example in our case by computing Poincar\'e
sections.

In fact, we showed that chaos arises precisely when exponential
amplification occurs, showing that the real source for these
amplifications is not a particular resonance condition, but it is
a consequence of the dynamical chaos of the background fields.

The evidence for an increasingly destruction of the invariant tori
shows, together with instability analysis, the existence of an
energy threshold above which the dynamics becomes chaotic, and the
amplifications become important.

We also address the subtle issue concerning the precise
relationship between chaos and parametric resonance. In this
context we showed a relationship between the metric entropy, which
is a measure of chaos, and the thermodynamic entropy, computed in
the high squeezing limit, which is also another way to see the
equivalence between the exponents. It may seem natural that both
FE and LE are equal, since both show exponential behavior of the
system. Nevertheless, the authors of Ref. \cite{feldkof} take the
exponential  behavior as ``rather formal'', and, being interested
only in the turbulent phase {\it after} preheating, define a
normalized distance in the phase space
\begin{equation}
\Delta (t)=\sum_{a}\left( \frac{f_{a}^{\prime
}-f_{a}}{f_{a}^{\prime
 }+f_{a}}
\right) ^{2}+\left( \frac{\dot{f}_{a}^{\prime
}-\dot{f}_{a}}{\dot{f} _{a}^{\prime }+\dot{f}_{a}}\right) ^{2}\;,
\end{equation}
according to which chaos sets in only when the preheating period
is over. The fundamental difference between this formula and the
one we used \cite{wolf,holden} is the normalization factors in the
denominators.  However, as we showed in Section \ref{model}, the
parametric resonance makes chaotic from the very beginning, by
construction. Of course, we restrict ourselves to the preheating
period. Therefore, our results do not concern the turbulent phase,
i.e. interaction with other modes, which would account for the
thermalization process.

%%%%%%%%%%%%%%%%%%%%%%%%%%%%%%%%%%%%%%%%%%%%%%%%%%%%%%%%%%%%%%%%%%%%%%%%

\section*{Acknowledgments}

The authors wish to thank Robert Brandenberger for useful comments
and discussions.  SEJ and VHC thank the High Energy Theory Group at
Brown University  where this work started. SEJ acknowledges financial
support from CNPq and
 partial support from the U.S. Department of
Energy under the contract
 D E-FG02-91ER40688 - Task A. VHC is
supported by the project FONDECYT grant 3010017.

%%%%%%%%%%%%%%%%%%%%%%%%%%%%%%%%%%%%%%%%%%%%%%%%%%%%%%%%%%%%%%%%%%%%%%%%

\begin{figure}
\centering
\epsfig{file=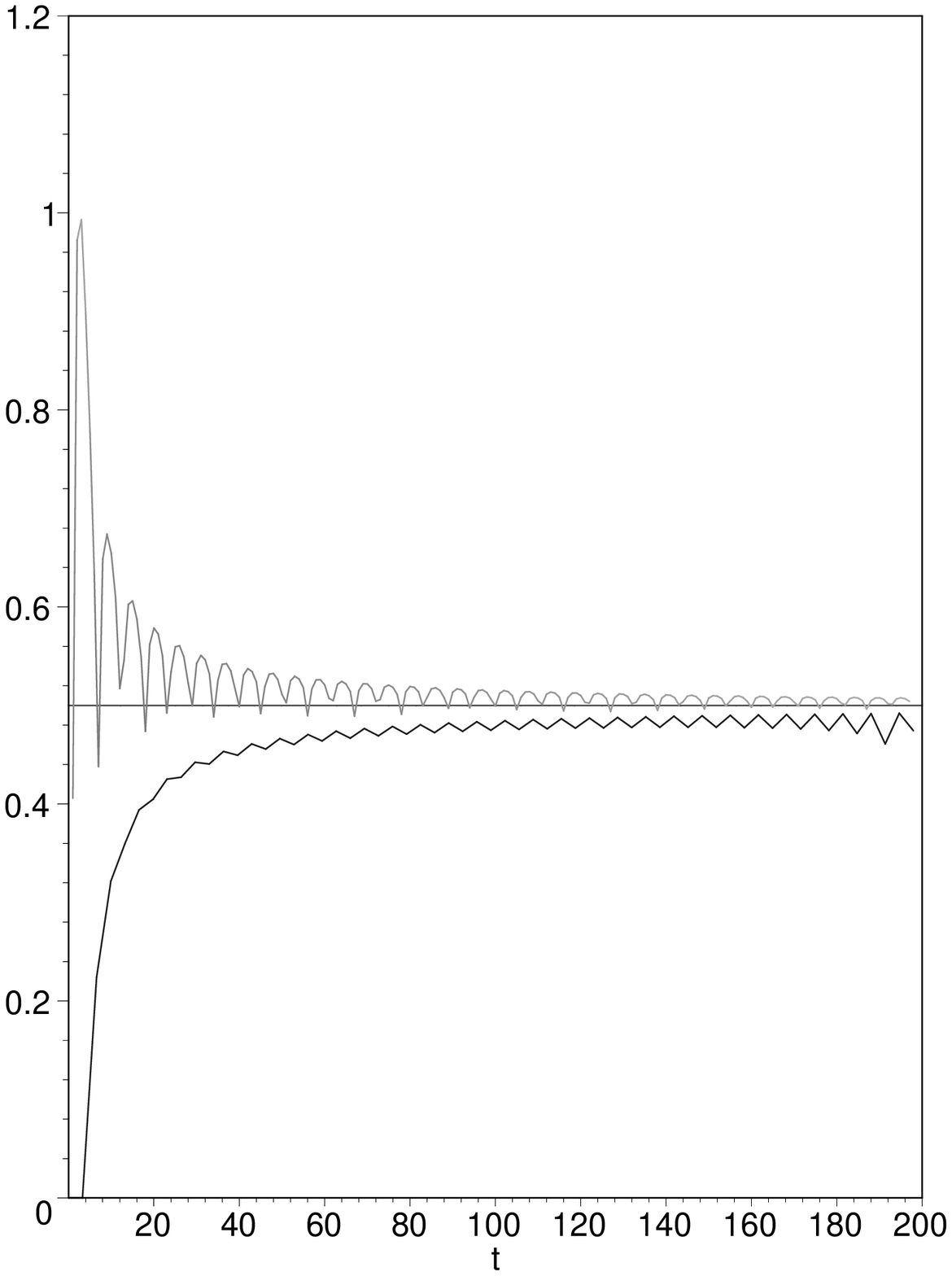} \caption{Plot of the LE for the exact
parametric resonance (upper curve). The straight line is the
theoretical value for the FE (for the parameters indicated in the
text); the lower curve is the FE as calculated by us.}
\label{parametric}
\end{figure}

\begin{figure}
\centering \epsfig{file=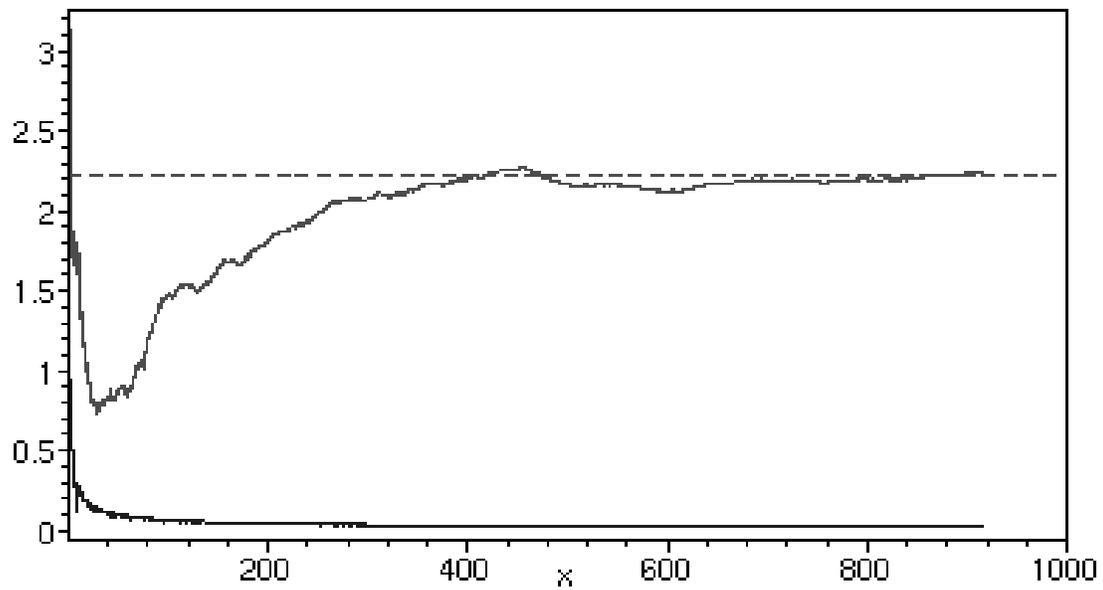,width=17cm}

\caption{Plot of the Lyapunov exponent for two different initial
conditions. For both trajectories, $M=10$, $m=1$ and $g=1$. The
dashed line
 indicates the limiting value for the non-vanishing LE.}
\label{le2}
\end{figure}

\begin{figure}[h]
\centering \epsfig{file=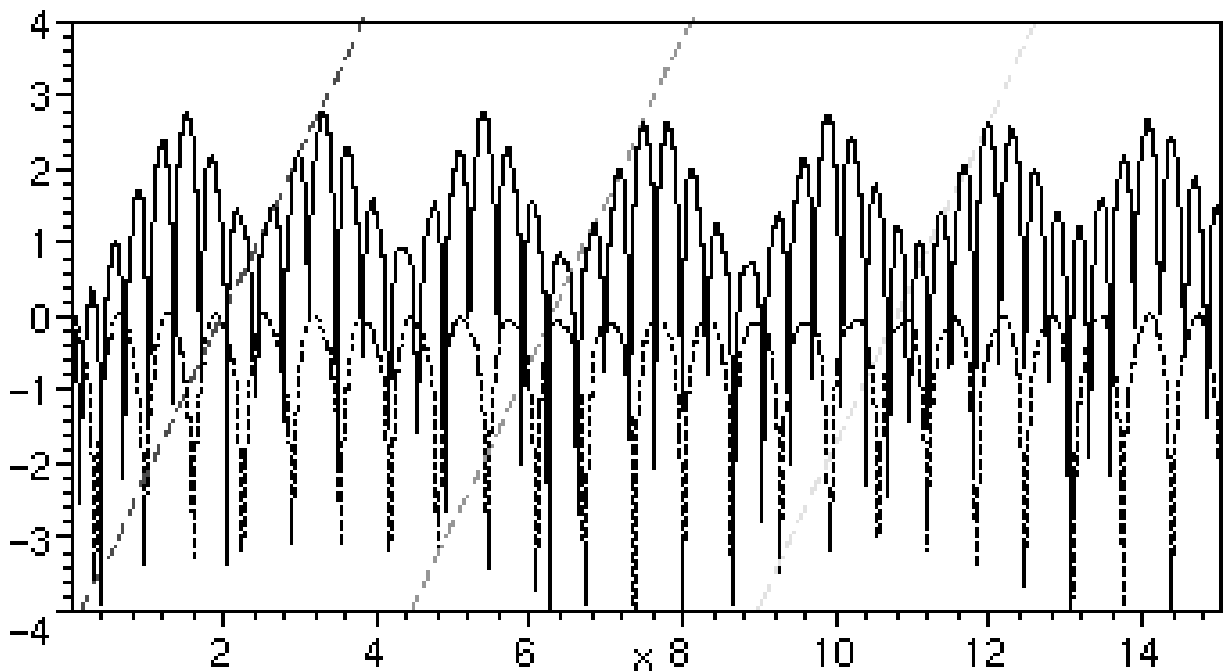,width=17cm}

\caption{Plot of $\ln(\protect\mu t) ~{\it versus}~   t$ for the
same
 initial  conditions used for the previous graph. The angular
coefficient of the straight dashed lines are given by the
non-vanishing LE from the  previous graph; their horizontal
positions are arbitrary.} \label{fe2}
\end{figure}

\begin{figure}[h]
\centering \epsfig{file=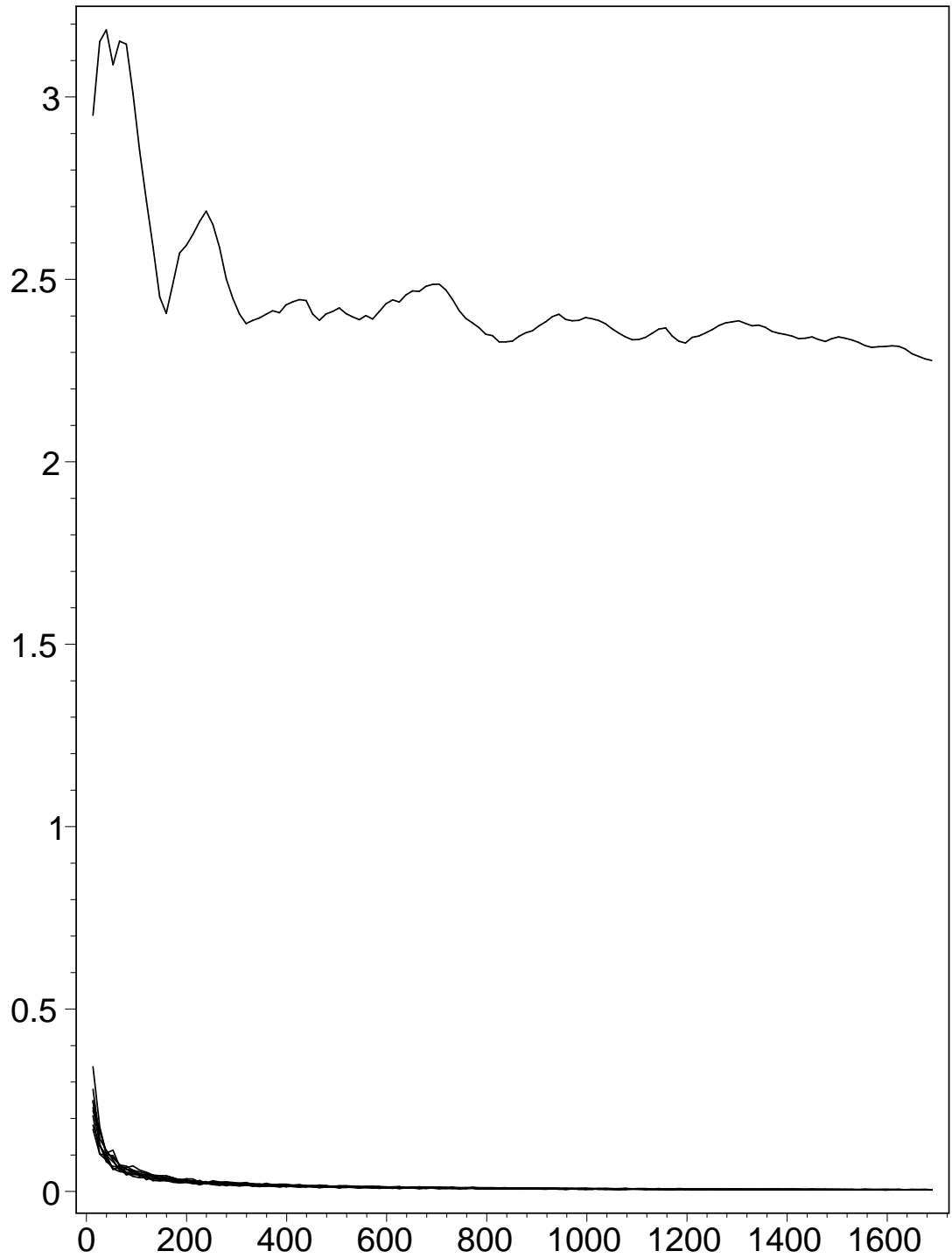,width=17cm}

\caption{Plot of the LE for ten different initial conditions. The
 upper line corresponds to a trajectory above a critical value for
the energy.} \label{thL}
\end{figure}

\begin{figure}[h]
\centering \epsfig{file=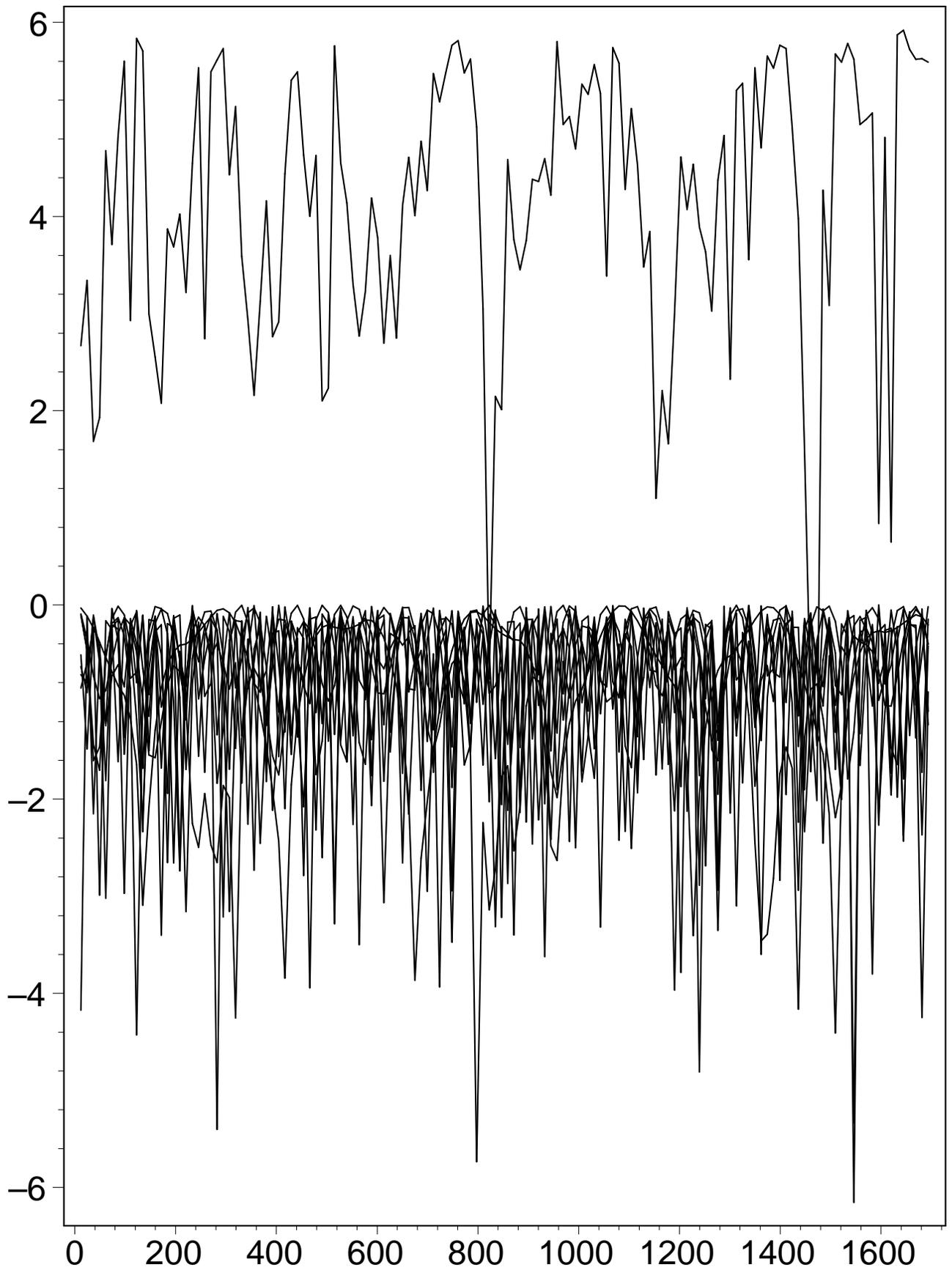,width=17cm}

\caption{Plot of the FE for the same initial conditions used in
the previous graph.} \label{thF}
\end{figure}

\begin{figure}
\centering
\epsfig{file=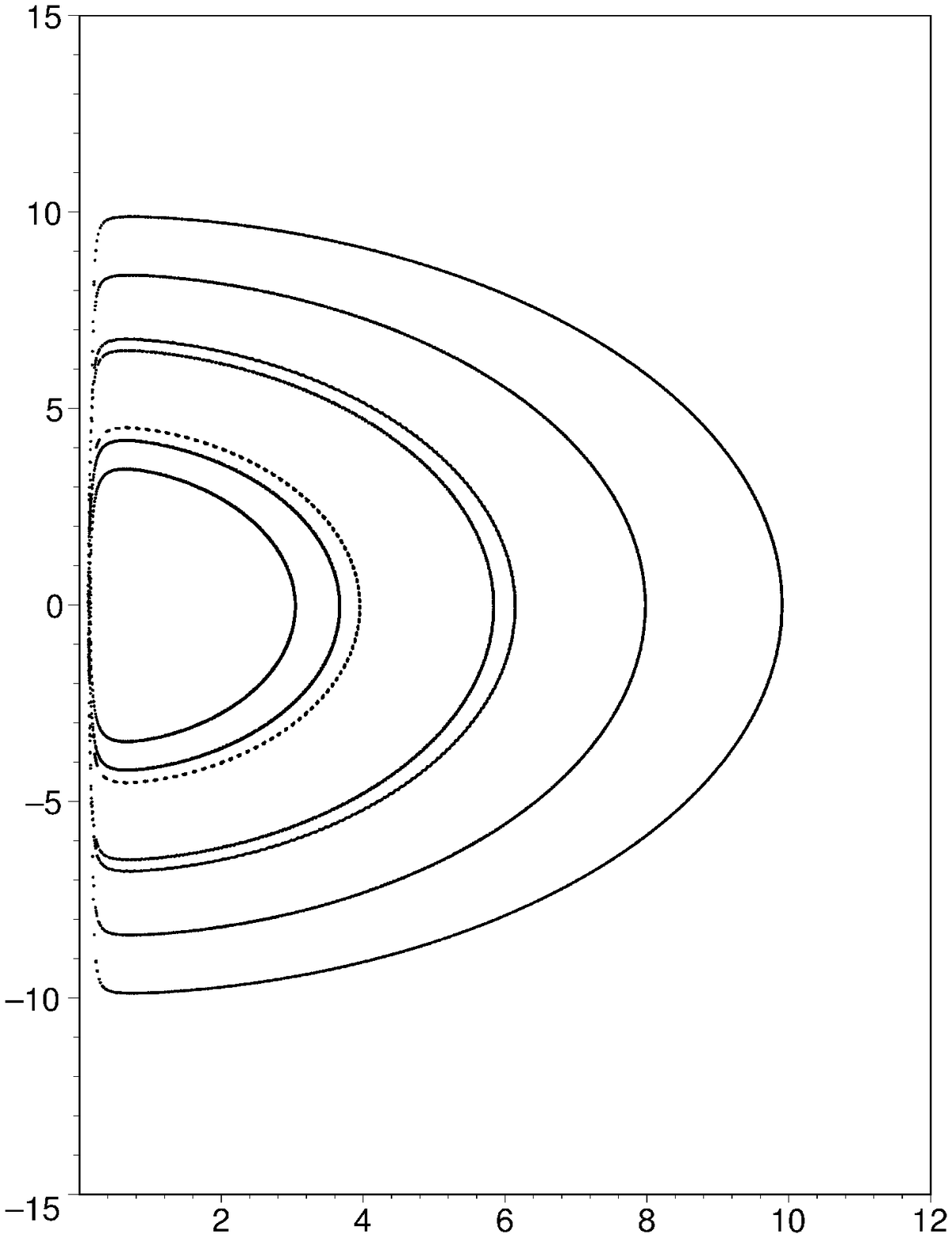,width=15cm,height=8cm}
\epsfig{file=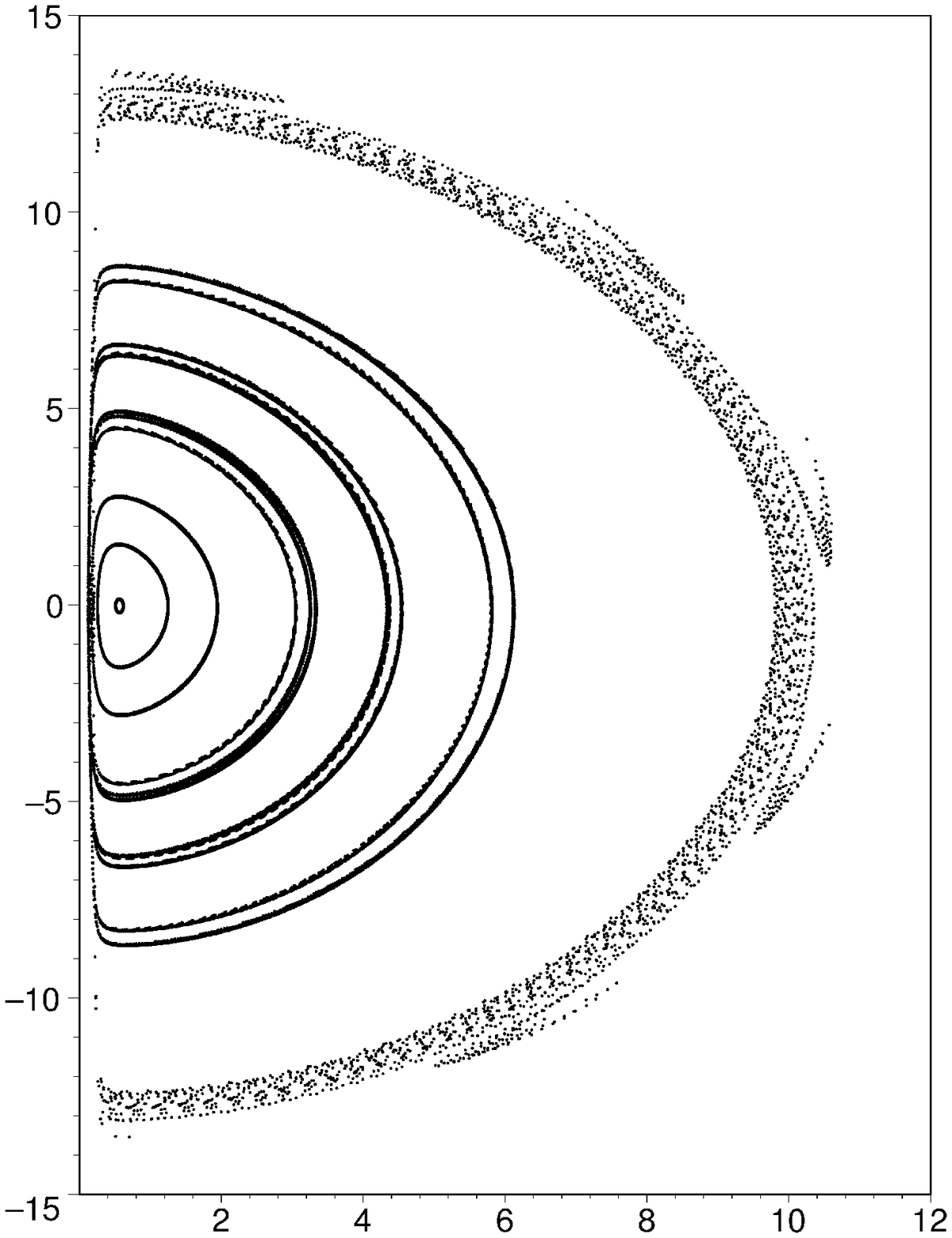,width=15cm,height=8cm}
\epsfig{file=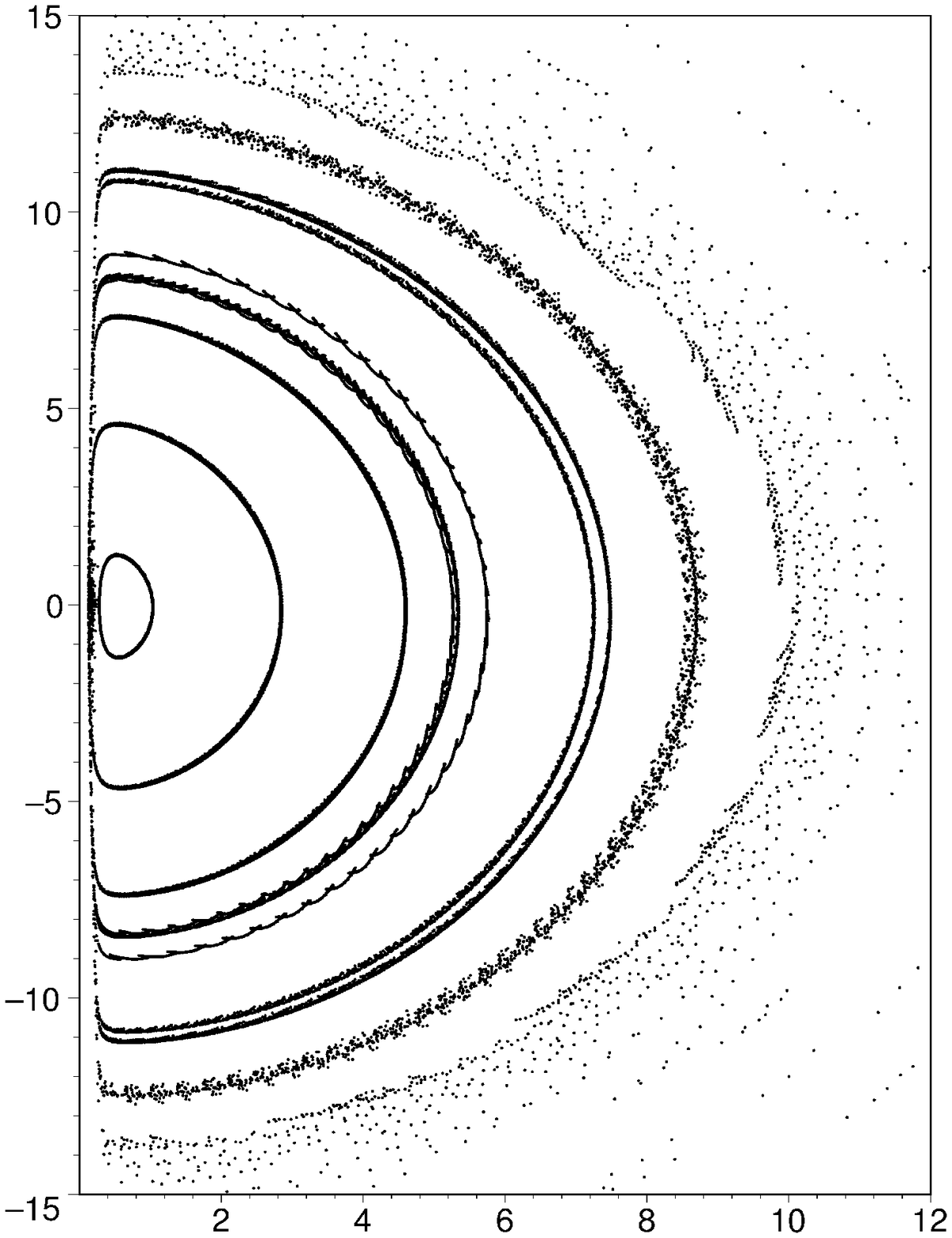,width=15cm,height=8cm}

\caption{Poincar\'e sections ($\dot R ~{\it versus}~  R$) for
 $E=50$,$150$ and $200$. One can clearly see the gradual
destruction of the tori as  the energy increases.}
\label{poincare}
\end{figure}

\end{document}